\documentclass[fleqn,10pt]{wlscirep}
\usepackage[utf8]{inputenc}
\usepackage[T1]{fontenc}
\usepackage{hyperref}
\usepackage{xcolor}

\title{Automated Cell Structure Extraction for 3D Electron Microscopy by Deep Learning}

\author[1]{Jin Kousaka}
\author[2,3,4]{Atsuko H. Iwane}
\author[1,2,3,*]{Yuichi Togashi}
\affil[1]{Graduate School of Life Sciences, Ritsumeikan University, 1-1-1 Noji-higashi, Kusatsu, Shiga 525-8577, Japan}
\affil[2]{Laboratory for Cell Field Structure, RIKEN Center for Biosystems Dynamics Research, 3-10-23 Kagamiyama, Higashi-Hiroshima, Hiroshima 739-0046, Japan}
\affil[3]{Laboratory for Comprehensive Bioimaging, RIKEN Center for Biosystems Dynamics Research, 2-2-3 Minatojima-minamimachi, Chuo, Kobe, Hyogo 650-0047, Japan}
\affil[4]{Research Institute for Cell Design Medical Science, Yamaguchi University, 1-1-1 Minami-Kogushi, Ube, Yamaguchi 755-8505, Japan}

\affil[*]{togashi@fc.ritsumei.ac.jp}


\begin{abstract}
Modeling the 3D structures of cells and tissues is crucial in biology. 
Sequential cross-sectional images from electron microscopy provide high-resolution intracellular structure information. 
The segmentation of complex cell structures remains a laborious manual task for experts, demanding time and effort. 
This bottleneck in analyzing biological images requires efficient and automated solutions. 

In this study, the deep learning-based automated segmentation of biological images was explored to enable accurate reconstruction of the 3D structures of cells and organelles. 
An analysis system for the cell images of {\em Cyanidioschyzon merolae}, a primitive unicellular red algae, was constructed.
This system utilizes sequential cross-sectional images captured by a focused ion beam scanning electron microscope (FIB-SEM). 
A U-Net was adopted and
training was performed to identify and segment cell organelles from single-cell images. 
In addition, the segment anything model (SAM) and 3D watershed algorithm were employed to extract individual 3D images of each cell from large-scale microscope images containing numerous cells. 
Finally, the
trained  U-Net was applied to segment each structure within these 3D images. 
Through this procedure, the creation of 3D cell models could be fully automated. 
The adoption of other deep learning techniques and combinations of image processing methods will also be explored to enhance the segmentation accuracy further. 
\end{abstract}

\begin{document}

\flushbottom
\maketitle
\thispagestyle{empty}

\section*{Introduction}

Advanced deep learning technology has significantly improved the accuracy and efficiency of general image recognition tasks. 
Furthermore, its application to biological tissue and medical imaging has attracted much attention.
As a demonstration, we analyzed 3D cell images captured using a focused ion beam scanning electron microscope (FIB-SEM). 
FIB-SEMs are primarily used to measure physical properties and evaluate microstructures of solid samples \cite{pmid11207928}. 
Recently, they have also been utilized to analyze 3D structures of cells and tissues.
The surface of resin-embedded biological samples is repeatedly cut using a gallium ion beam and imaged by a scanning electron microscope. 
The serial block-face (SBF) image data, i.e., a sequence of cross-sectional images, is reconstructed as a 3D structure. 

However, this technology still requires improvement.
The intricate boundaries separating structures within each slice, low contrast, and significant noise make analysis based on brightness challenging. 
Furthermore, as images taken with an electron microscope are single channel (grayscale), relying on electron beam reflections, the absence of color information further complicates the analysis.
In addition, identifying specific cells within a space containing multiple cells and extracting their microstructures is difficult. 
Consequently, the segmentation of microstructure depends on subjective evaluation and manual techniques performed by experts, consuming a considerable amount of time. 
Image data analysis, which automates this process and is more efficient, objective, and quantitative, is expected to contribute to the analysis of various 3D biological images.
The objective of this study was to automate this process fully, exploring the application of deep learning in the recognition tasks of cultured cell images. 

\begin{figure}[h]
\centering
\includegraphics[width=\linewidth]{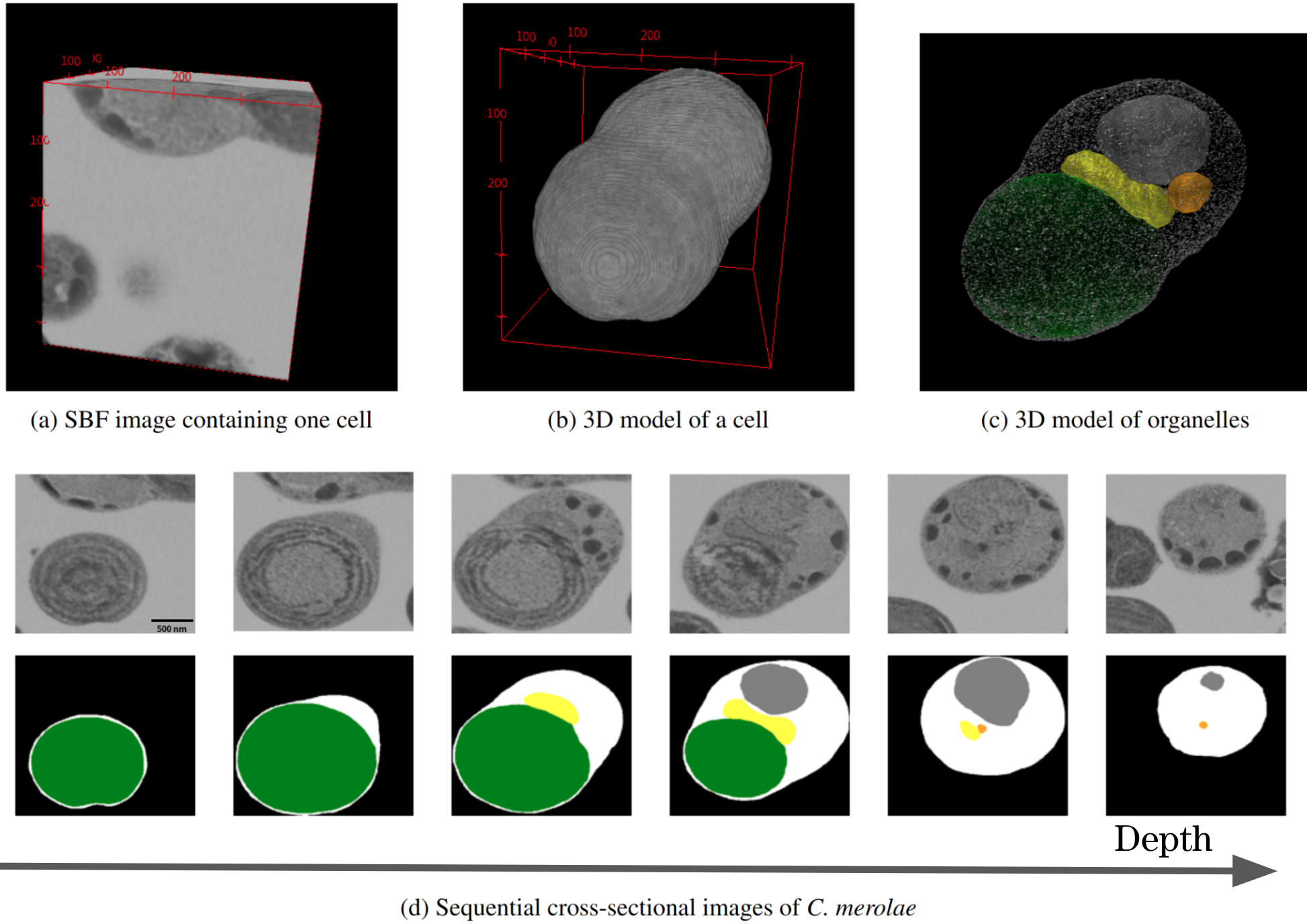} 
\caption{Procedure to reconstruct a 3D model of a cell from sequential cross-sectional images. (a) SBF image that centrally includes the cell of interest, cropped from FIB-SEM images. 
(b) By reconstructing the data and removing the background, a 3D model of the cell is produced. 
(c) Three-dimensional organelle model visualized by further removing the cytoplasm, indicating cytoplasm (white), plastid (green), mitochondrion (yellow), peroxisome (orange), and nucleus (gray).
(d) This SBF consists of sequential cross-sectional images, as depicted in the top row. Various organelles, the cytoplasm, and the background in these images were manually identified by experts, as shown in the bottom row. 
In this panel, images at six different depths (in the direction perpendicular to the images) are presented.}
\label{figure:figure1}
\end{figure}

\begin{figure}[ht]
\centering
\includegraphics[width=\linewidth]{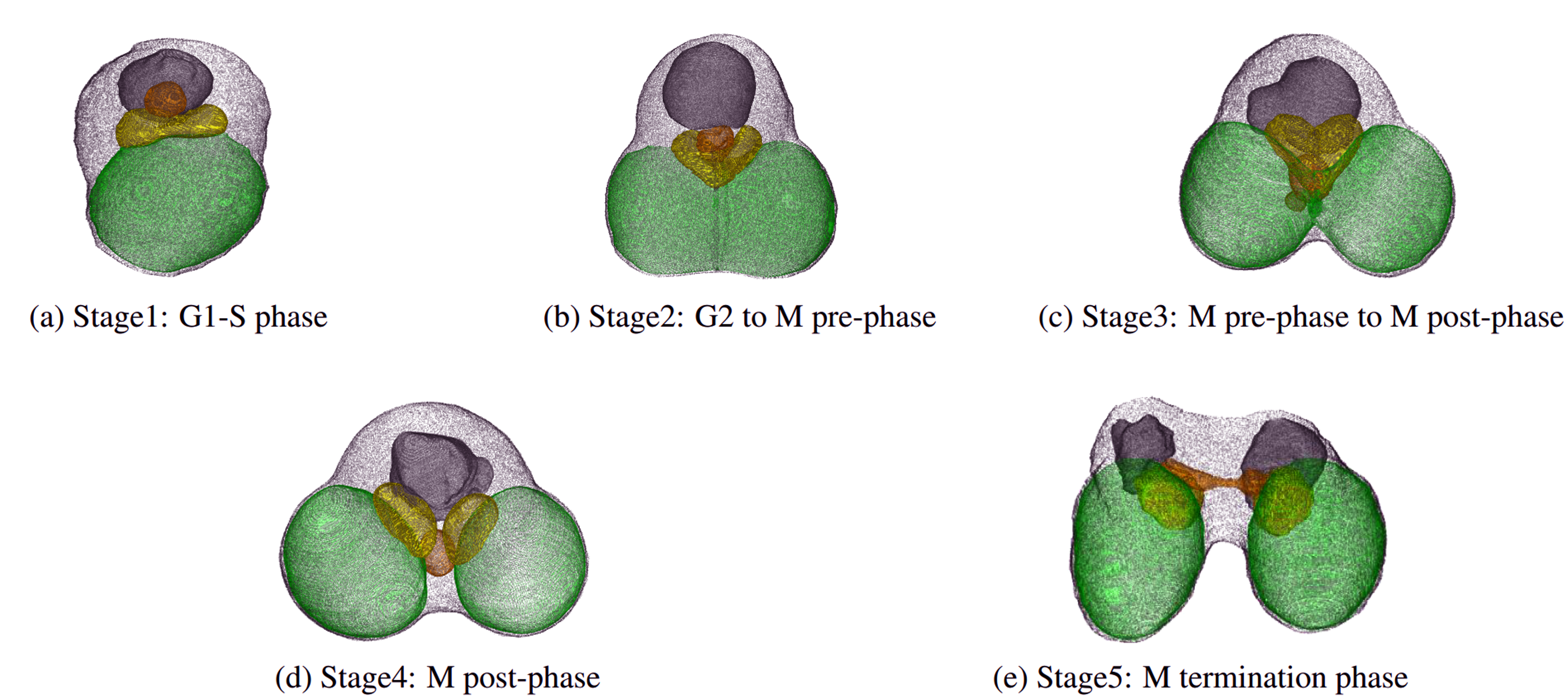}
\caption{Structural changes associated with the cell cycle of {\it C. merolae}. Each structure includes the cytoplasm (white), plastid (green), mitochondrion (yellow), peroxisome (orange), and nucleus (gray). Each type of organelle is equally distributed to the daughter cells during division  \cite{pmid9522454,pmid15681662}. The division of organelle structures occurs at nearly consistent timings regardless of the individual, although the timing varies depending on the type of organelle \cite{pmid33580410}.
}
\label{figure:figure2}
\end{figure}

We chose {\it Cyanidioschyzon merolae} as a model for the automated reconstruction of 3D cell structures using an FIB-SEM \cite{jin_sci,Takasaka2023}. 
{\it C. merolae} is a primitive unicellular eukaryote belonging to red algae, with a diameter of approximately 1.5--2.0 $\mu$m \cite{1010000781756803720} (Figure \ref{figure:figure1}).
This organism has a simple structure with only one organelle of each type: nucleus, mitochondrion, plastid, and peroxisome \cite{pmid33836072}. 
Each organelle is distributed equally to the two daughter cells \cite{pmid9522454,pmid15681662} (Figure \ref{figure:figure2}).
The genome of {\it C. merolae} and the genomes of its mitochondrion and plastid have been completely sequenced \cite{pmid9801318,pmid12755171,pmid15071595,pmid17623057}. 
Moreover, it is relatively easy to synchronize the cell division cycles of multiple {\it C. merolae} cells in culture with the cycle of light \cite{pmid8082652}. 
Due to these characteristics, {\it C. merolae} is regarded as a model organism for investigating the mechanisms of cell division. 
Cell division in {\it C. merolae} has been extensively studied using molecular and cell biological approaches.
Recent advances in observation techniques have enabled the observation of dynamic changes in the behavior and target molecules of individual living cells \cite{pmid33580410}.
However, the physical mechanism of division control based on mechanical properties remains unclear. 

Therefore, this article proposes an automated method for accurately extracting cell and organelle structures during the division process of {\it C. merolae}. 
In previous studies, a technique for segmenting each structure from {\it C. merolae} FIB-SEM images using machine learning was proposed \cite{pmid32060319}. 
However, this method segments structures from each cross-sectional image in the SBF image containing only one cell. 
Separating cells and backgrounds is still performed manually and is time-consuming. 
Additionally, the method disregards the sequential contours between cross-section images and has problems in the segmentation of structures at the cell edge. 
This article proposes a technique to resolve such shortcomings. 
This method, which enables the automated and rapid acquisition of 3D structures of multiple {\it C. merolae} cells, will facilitate the investigation of the mechanical characteristics related to cell division in the future.

\section*{Materials and Methods}

\subsection*{Image Acquisition and Preprocessing}

\subsubsection*{Image data acquisition using FIB-SEM}

The images of samples containing multiple {\it C. merolae} cells, physically fixed by high-pressure freezing, were obtained by an FIB-SEM.
The aspect ratios of the obtained sequential sectional images were corrected, and the noise was removed using a median filter using Amira software (Thermo Fisher Scientific Inc.). 
Experts performed cropping to construct learning data images, to create SBF images each containing one cell with side lengths of approximately 200-500 pixels, as depicted in Figure \ref{figure:figure1} (a). 
The experts manually segmented each sequential cross-sectional image constituting the SBF image to label the plastid, mitochondrion, peroxisome, nuclei, cytoplasm, and background \cite{pmid32060319}. 
In other words, all voxels constituting the SBF images were labeled into six classes. 

The cell cycle of {\it C. merolae} is classified into five stages \cite{pmid8082652}, shown in Figure \ref{figure:figure2}. 
The G1-S phase of the cell cycle is referred to as Stage 1, followed by the G2 phase to the M phase (pre-phase) as Stage 2, the M phase (pre-phase) to the M phase (post-phase) as Stage 3, the M phase (post-phase) as Stage 4, and the M phase (termination phase) as Stage 5. Multiple SBF images were labeled for each stage.
The labeled SBF images totaled 68 and were classified into one of these stages by experts.

\subsubsection*{Image data preprocessing}

The images obtained by FIB-SEM and the labeled dataset created by experts were preprocessed for deep learning analysis; due to a difference in resolution between the horizontal and vertical directions when capturing images with the FIB-SEM, the voxels in the SBF images were shaped as rectangular parallelepipeds rather than cubes. 
The SBF images were resized using nearest-neighbor interpolation to restore the aspect ratios of the voxels to their actual size ratios, facilitating 3D data analysis. 
A bilateral filter technique was employed to process the images and remove noise that Amira software could not eliminate \cite{BCCS12}. 
Each sectional image was resized to create cubic voxels, each side measuring 256 pixels, to ensure a consistent input image size for the deep learning model. 
For effective deep learning, the brightness values of the electron microscope images were normalized to a range of 0--1 (by dividing 8-bit values by 255). 
After these preprocessing steps, the dataset was split into different sets for cross-validation, resulting in six cases. 
In each case, an SBF image was selected as validation data to
train the deep learning model, each corresponding to a different stage of cell division. 
The remaining SBF images were used as training data for the deep-learning model. 
In addition to the conventional cross-sectional images obtained from the top surface by FIB shaving, the SBF training image dataset was augmented by creating new cross-sectional images from both lateral and anterior views. 
Details on the number of cells used and the number of cross-sectional images for each dataset in each cross-validation case are presented in Table \ref{tab:table1}.

\subsection*{Automated Segmentation of Single Cells Using Deep Learning}

\subsubsection*{Deep neural network architecture}

\begin{figure}[ht]
\centering
\includegraphics[width=\linewidth]{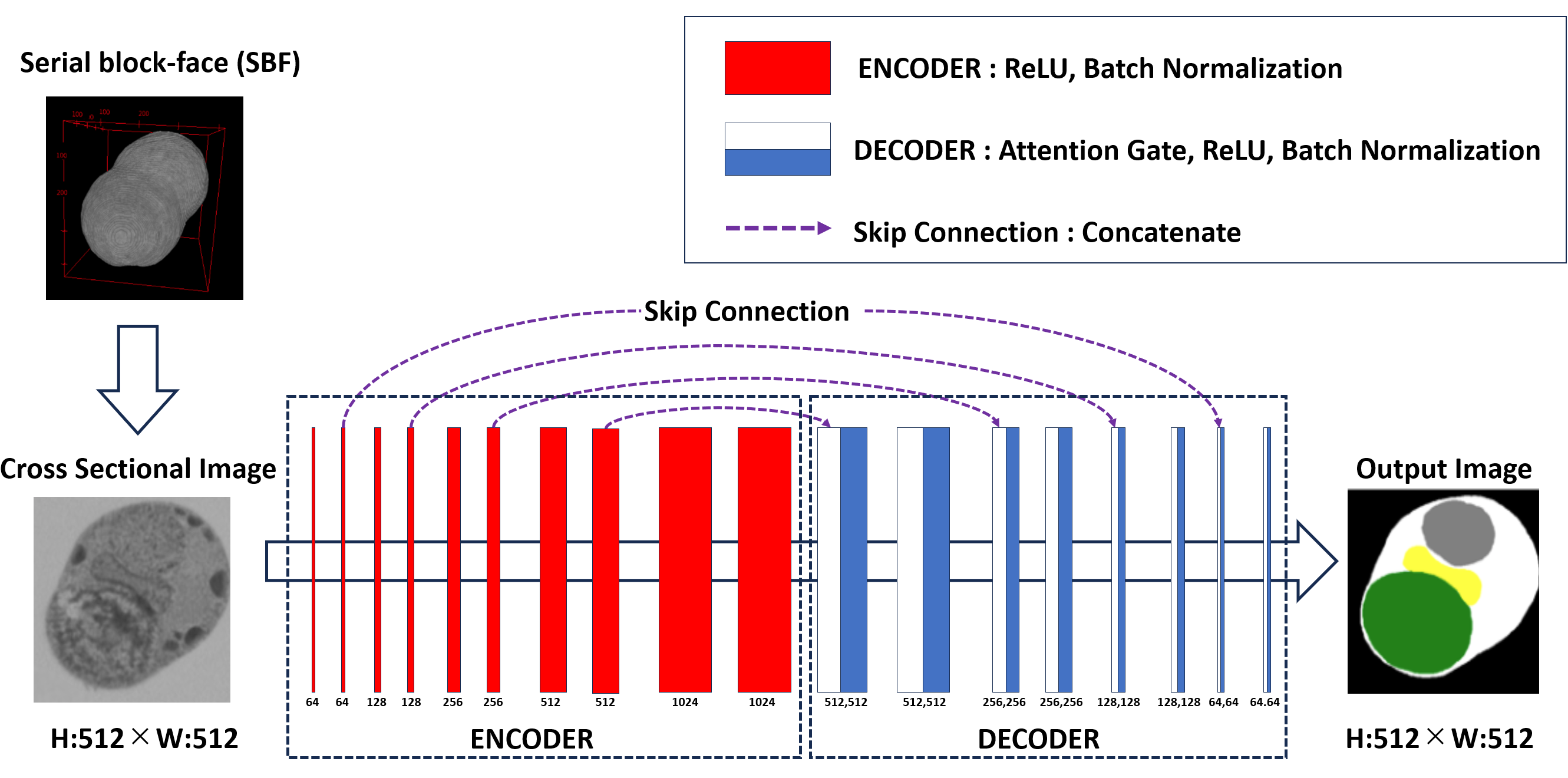}
\caption{The architecture of U-Net (W and H are output width and height, respectively).}
\label{figure:figure3}
\end{figure}

We developed a scanning 3D U-Net neural network model, as illustrated in Figure \ref{figure:figure3} (parameters are shown in Table \ref{tab:table2}) 
{\cite{NIPS2012_c399862d,ioffe2015batch,cicek2016_3dunet}}. 
Initially created for image segmentation tasks, U-Net is now extensively employed in various fields, including medical imaging \cite{10.1007/978-3-319-24574-4_28}. 
The first half of the neural network consists entirely of convolutional layers. 
On the opposite side, deconvolutional layers gradually restore the image size to the original input. 
The convolutional and deconvolutional layers are connected by skip connections, designed to prevent information loss due to the encoding of the convolutional layers. 
In this study, we implemented four stages of convolution and deconvolution, with each deconvolution layer in our model enhanced with an attention gate mechanism \cite{vaswani2023attention}. 
The ReLU function was used as the activation function at each layer, and batch normalization was performed at all layers to prevent overfitting \cite{NIPS2012_c399862d,ioffe2015batch}. 
The U-Net model in this study had approximately 130 million parameters.
We performed a six-class segmentation task. 
The input data shape was defined as a tensor of [batch size, number of channels, vertical pixel count of the SBF image, and horizontal pixel count of the SBF image]. 
The label data and U-Net output values were fed into the TverskyLoss function for loss score calculation \cite{DBLP:journals/corr/abs-1810-07842}. 
The AdamW learning function was used, with the learning rate starting from a base rate of 0.00001 and decaying by a factor of 0.95 each epoch \cite{DBLP:journals/corr/abs-1711-05101}. 
The number of epochs was set to 5. 
All networks were built on a GPU (NVIDIA GeForce RTX 4090 with 24 GB memory), and the framework was implemented using Python and TensorFlow. 
To accelerate learning, the tensor cores of the GPU  were automatically utilized.

\subsubsection*{Segmentation accuracy}

During the
training phase, we evaluated the accuracy of the learning model through cross-validation. 
Our cross-validation process included six distinct cases (Table \ref{tab:table1}); for each case, a dataset of 63 SBF images was used for training, and a separate set of 5 SBF images (one for each stage in the cell cycle) was employed for validation. 
The SBF images for validation did not overlap between cases.
Each validation set included one cell from each of the five stages of the cell division cycle. 
We segmented the sequential cross-sectional images into six classes using the U-Net for each case. 
The output of the U-Net approached 1 if the segmentation results indicated a specific class and approached 0 otherwise. 
Finally, we compared these inference results with the label data and evaluated the model using the intersection over union (IoU) as expressed:
\begin{equation}
    \label{math:math1}
IoU = \frac{\text{Area of overlap}}{\text{Area of union}}.
\end{equation}
The area of overlap is the intersection of the predicted and ground truth areas,
and the area of union refers to the union of the predicted and ground truth sets.

\subsection*{Automated Extraction of Individual Cells from Large-Scale Images}


In this study, we first implemented U-Net-based automated segmentation of organelles using single-cell SBF images that were manually annotated by experts. 
However, because a model trained solely on single-cell data may have challenges accurately delineating boundaries in large-scale images containing multiple cells, a complementary approach for extracting individual cells from such images is required.

\subsubsection*{Segment anything model}

To generate SBFs automatically, we initially utilized the segment anything model (SAM) to distinguish cells and backgrounds automatically. 
Developed by Meta, the SAM enables easy segmentation of cells and backgrounds from large-scale microscopic images containing multiple cells \cite{kirillov2023segment} 
({Although SAM2 has recently been announced, this study utilizes SAM, as that was the version available at the time of the research}). 
SAM was trained on over 110 million images using over one billion masks and possesses zero-shot capabilities, enabling it to extract structures from unknown classes. 
We employed SAM to extract cells from the background in microscopic images containing multiple cells.
Although SAM demonstrates high accuracy in cases with clear cell contours, it is less effective in instances in which cell contours are less defined, as shown in Figure \ref{figure:figure6}. 
This situation presents challenges in extracting cellular organelles from SBFs with indistinct cell outlines. 
Therefore, although SAM works well for separating cells from backgrounds, a U-Net model trained to recognize cell organelles from SBF images is needed to overcome its limitations in more complex images.

\subsubsection*{Three-dimensional watershed algorithm}

The next challenge was automating individual cell identification after completing the classification between cells and backgrounds using SAM segmentation. 
For this task, we employed the 3D watershed method. The watershed algorithm segments image regions based on brightness values \cite{87344}. 
This algorithm simulates water flowing from mountains to valleys, dividing the image into different areas. 
The 3D watershed algorithm extends this concept to three dimensions, enabling the segmentation of the image space by structural region. 
This method enables the identification of structures on a per-cell basis. Typically, images are binarized into black and white before implementing this method, clearly distinguishing between areas of high and low brightness values. 
A specific brightness threshold is needed for this binarization.

Applying the watershed method directly to noisy images can be challenging, especially when noise obscures the contours of structures, hindering effective segmentation. 
To address this issue, we initially performed SAM segmentation for large-scale microscopic images containing multiple cells. 
This preliminary step facilitated image binarization by distinguishing between cells and background. Although SAM operates on 2D images and cannot fully determine the continuity between structures in consecutive images, it significantly aids initial segmentation.
Following SAM analysis, we applied the 3D watershed algorithm for further segmentation, making SAM a supportive tool in the 3D watershed process.

\section*{Results}

\subsection*{Performance of U-Net Segmentation on SBF Images}

\begin{figure}[ht]
\centering
\includegraphics[width=\linewidth]{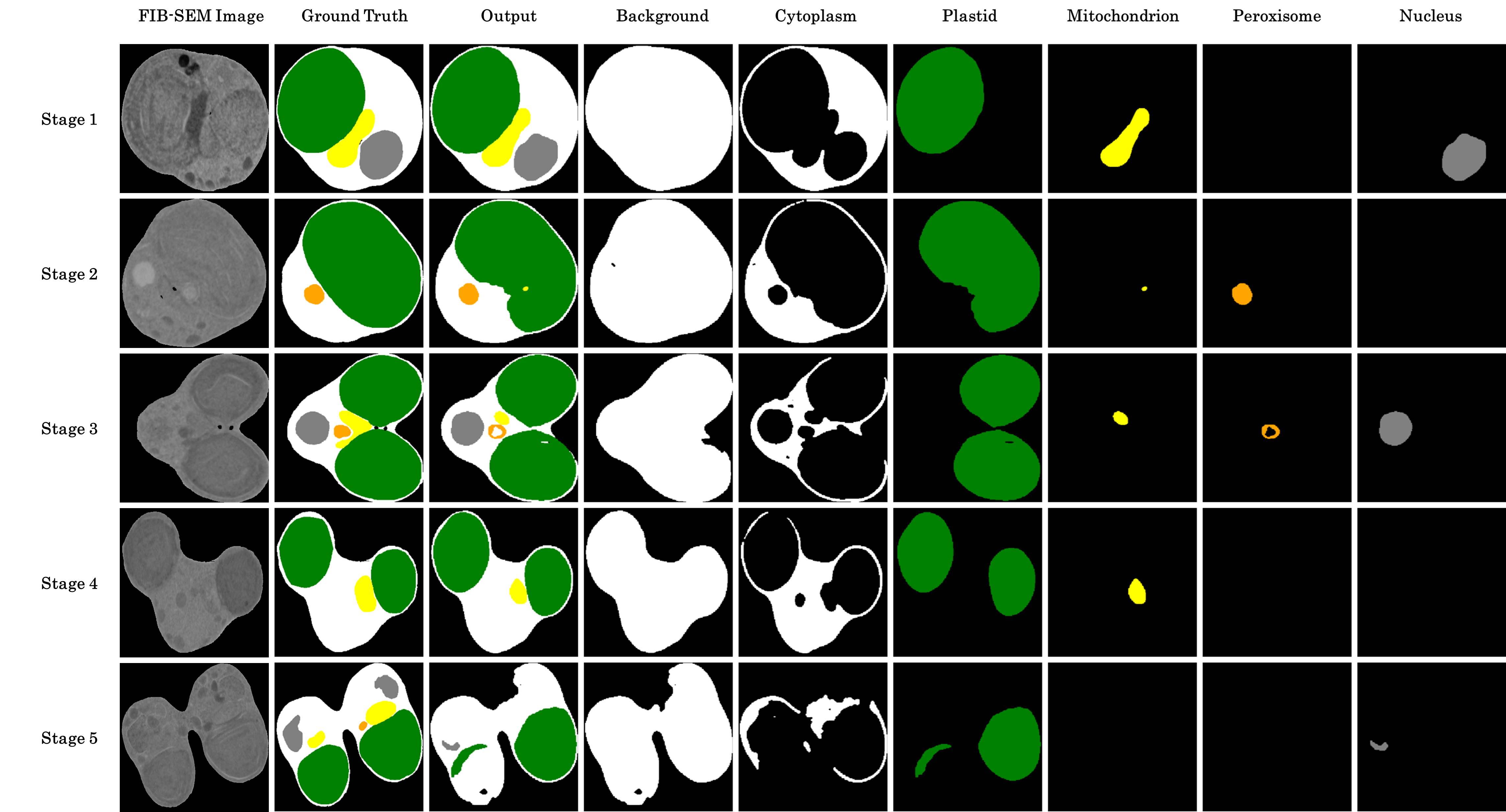}
\caption{Examples of
inference results for the validation dataset using a model obtained through the
training. Each structure includes the cytoplasm (white), plastid (green), mitochondrion (yellow), peroxisome (orange), and nucleus (gray). The output images of the plastid and cytoplasm closely match the ground truth.}
\label{figure:figure4}
\end{figure}

Table \ref{tab:table3} presents the IoU score for each case based on the inference results of the U-Net. 
The table organizes six training classes by row and five cell division stages by column. 
These results indicate that the average accuracy of background identification is approximately 0.99, suggesting that cells and backgrounds can be identified with very high accuracy, as illustrated in Figure \ref{figure:figure4}. 
Additionally, the accuracy of plastid inference remains high throughout all stages of cell division. 
In contrast, the mitochondrion and nuclei are smaller in volume than the plastid, making them more difficult to identify accurately, with an average accuracy of about 0.70. 
The average accuracy for peroxisomes is notably lower, approximately 0.45.
Furthermore, the inference accuracy for all classes tends to decrease gradually as the cell division stage progresses. 

Plastids are relatively easily identifiable due to their size, unique structure, and distinct brightness values. 
In contrast, mitochondria and nuclei have brightness values similar to those of the background and starch, making them more challenging to identify. 
Additionally, peroxisomes are assumed to be difficult for deep learning models to learn due to their irregular shapes that vary from cell to cell and brightness values that are similar to those of the background and starch \cite{miyagishima1999,ijms21155452}. 
Furthermore, as cell division progresses, especially before division, a single SBF image may encompass two cells, increasing the complexity of cell structures within SBF images and potentially decreasing the inference accuracy as cell division advances.

\subsection*{Multi-Directional Inference and Its Impact
}

\begin{figure}[ht]
\centering
\includegraphics[width=\linewidth]{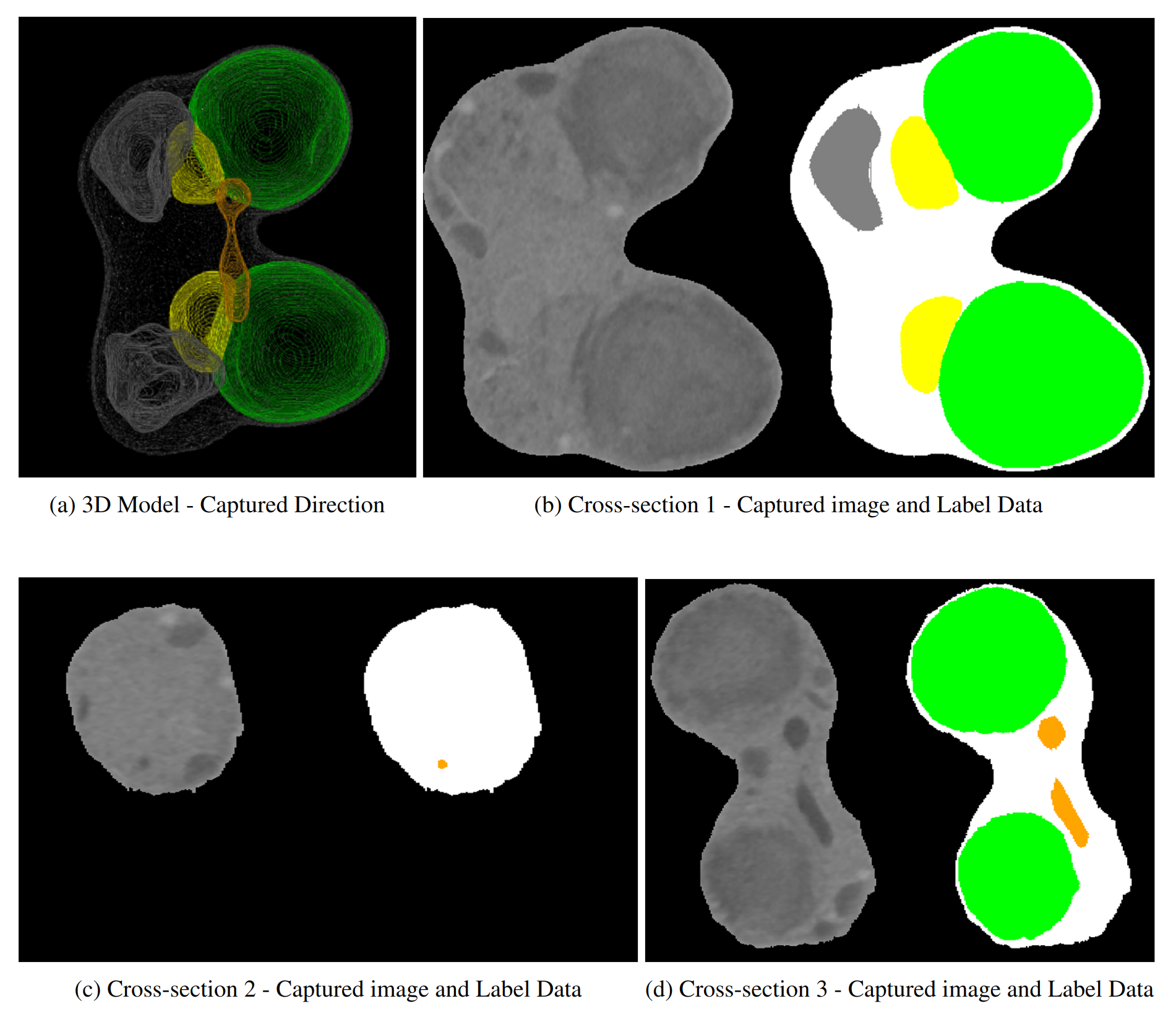}
\caption{Cross-sections and label data of cells in various directions. (a) Three-dimensional cell model reconstructed from captured consecutive cross-sectional images. This model is visualized from the captured direction. (b) Consecutive cross-sectional images taken near the center of the cell (left) and label data created by experts based on them (right). (c) and (d) New cross-sectional images generated from stacked consecutive cross-sectional images (left) and the corresponding label data (right). These images were generated from two directions, one aligned with the captured direction and the other perpendicular. Each panel includes the cytoplasm (white), plastid (green), mitochondrion (yellow), peroxisome (orange), and nucleus (gray).}
\label{figure:figure5}
\end{figure}
\begin{figure}[ht]
\centering
\includegraphics[width=\linewidth]{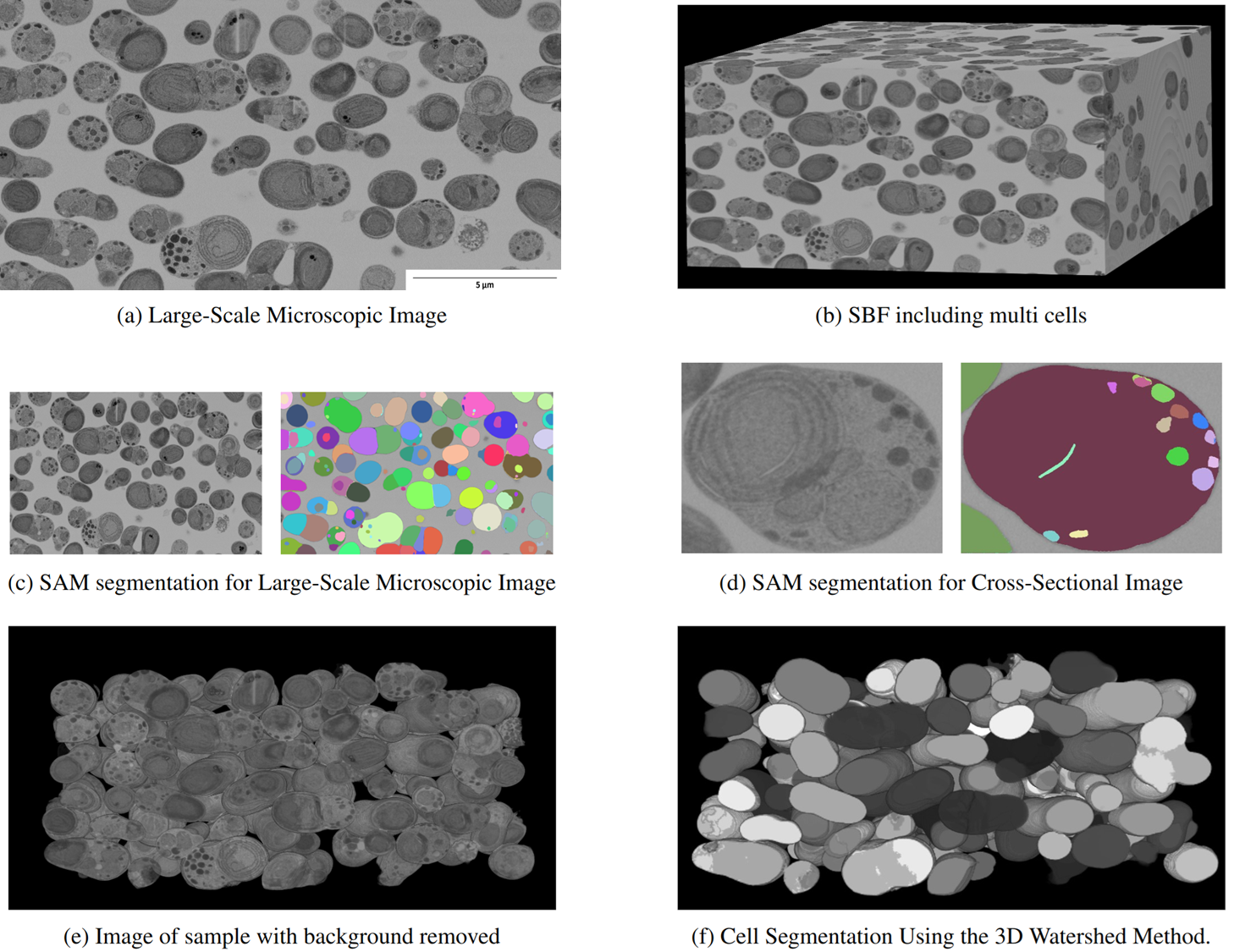}
\caption{Schematic diagram of automatic identification of individual cells from a 3D image containing multiple cells using SAM and the 3D watershed method. (a), (b) Cross-sectional image of a sample containing multiple cells and its SBF. (c) Extraction of cells from the background by SAM. (d) Classification of cellular organelles by SAM. (e) Removal of the background identified by SAM from the SBF in (b). (f) Identification and color-coding of individual cells by the 3D watershed method.
}
\label{figure:figure6}
\end{figure}

We segmented various cellular organelles from cross-sectional images of 2D pictures.
However, a problem was identified with low identification accuracy in some classes. 
To address this issue, we generated new cross-sectional images perpendicular to the original images, using the 3D SBF data, as shown in Figure \ref{figure:figure5}.
We used these images for inference with a
trained model. 
This model calculates the probability of belonging to all classes for each voxel (the smallest unit of a 3D image). 
Three predictive values are obtained for each voxel with inference from three directions. 
The class with the highest probability is selected when different classes are predicted. 
As this process may introduce noise, a median filter removes noise after selection. 
Consequently, as shown in Table \ref{tab:table3}, some cases exhibit no change or a slight decrease in predictive accuracy, but other samples show clear improvements in accuracy. 
This finding suggests that accurately determining the class from a cross-sectional image from a specific direction may not be possible. 
Also, considering that the orientation of the cell to be imaged cannot be predetermined and that inference from difficult angles is inevitable, inferring from three directions is worthwhile, even if it means accepting a slight decrease in accuracy.

\subsection*{Application to Large-Scale Microscopic Image Segmentation}

We used an ImageJ plugin to process the continuous cross-sectional images captured by the FIB-SEM and applied the 3D watershed algorithm.
This technique enabled us to identify different cells individually, as shown in Figure \ref{figure:figure6} (f). 
Thus, the process of creating an SBF image containing a single cell and identifying each cell and cell organelle, previously done manually by experts, was fully automated.

\section*{Discussion}

We successfully established a method of automatically segmenting and focusing on specific cells within a mixed spatial environment in 3D imagery.
By incorporating the latest deep learning models, which are being released at an ever-increasing rate, we can further enhance the accuracy of our inferences.
This research method was achieved by employing deep learning analysis techniques, which require a substantial amount of high-quality, expertly labeled data as a prerequisite. 
The necessity for extensive labeled data poses a potential bottleneck for future advancements in this field.
This research holds significant promise for biologists, particularly in aiding the morphological and dynamic analysis of cellular and tissue structures in 3D. 
The ability to segment and analyze intricate cellular compositions automatically in three dimensions provides a powerful tool for biological research.

{In principle, the technology presented here could be applied to images obtained by different types of 3D imaging devices, not limited to FIB-SEM. However, it appears especially effective to use machine learning for processing electron microscope images that contain only brightness information.
A large volume of high-quality annotated data corresponding to 3D images is required to utilize this technology. This poses the most significant challenge when applying the technology to other cell types.
Recently, various methods have been proposed for creating label data (semi-)manually and efficiently, and for performing zero-shot segmentation without label data. By integrating these methods, we may further enhance the technology described here to extract only specific structures from the environment crowded by multiple cells.
In addition, it was suggested that label data of diverse patterns of division would be required when performing structure extraction for various cell states such as cell division, where the determination of the cell state would be another target of machine learning.
}

\section*{Data availability}

The analysis codes and the raw data of the analysis results in this study are available from the corresponding author (Y. T.) on reasonable request.

\bibliography{main}

\section*{Acknowledgements}

This work was supported by JSPS KAKENHI grant number JP21K03487.

\section*{Author contributions statement}
J. K. conceived and implemented the analysis algorithms.
A. H. I. provided the 3D cell images.
Y. T. supervised the research.
All authors wrote and reviewed the manuscript.

\section*{Additional information}

The authors declare no competing interests.

\begin{table}[ht]
 \caption{Dataset used for cross-validation of 
training. ``Case Number'' refers to the consecutive number for each case of cross-validation. The number of cells used for training data in each cell cycle is denoted as ``Cells'', and the number of cross-sectional images used for training is indicated as ``LI''. ``VI'' represents the number of cross-sectional images used for the validation data.} 
 
  \centering
  \scalebox{0.70}[0.70]{ 
  \begin{tabular}{ccccccccccccccccccc}
    \toprule
     & \multicolumn{18}{c}{Cell Cycle Stage} \\
    \cmidrule(r){2-19}
    Case Number &\multicolumn{3}{c}{Stage1} &\multicolumn{3}{c}{Stage2} &\multicolumn{3}{c}{Stage3} &\multicolumn{3}{c}{Stage4}  &\multicolumn{3}{c}{Stage5} &\multicolumn{3}{c}{Total} \\
    \cmidrule(r){2-19}
    & Cells & LI & VI & Cells & LI & VI & Cells & LI & VI & Cells & LI & VI & Cells & LI & VI & Cells & LI & VI   \\

    \midrule\midrule
    1     & 24 & 16,568 & 482 & 12 & 15,701 & 588 & 11 & 13,852 & 426 & 11 & 15,098 & 550 & 5 & 7,695 & 424    & 63 & 68,914 &  2,470\\
    2     & 24  & 16,660 & 470 & 12  & 15,520 & 342 & 11  & 13,854 & 432 & 11  & 14,974 & 580 & 5  & 7,327 & 672    & 63 & 68,335 & 2,496 \\
    3     & 24  & 16,533 & 464 & 12  & 15,233 & 514 & 11  & 13,798 & 470 & 11  & 15,373 & 416 & 5  & 7,218 & 800    & 63 & 68,155 & 2,664 \\
    4     & 24  & 16,696 & 404 & 12  & 15,245 & 420 & 11  & 13,881 & 362 & 11  & 15,537 & 384 & 5  & 7,725 & 370    & 63 & 69,084 & 1,940 \\
    5     & 24  & 16,723 & 340 & 12  & 15,185 & 478 & 11  & 13,805 & 388 & 11  & 15,452 & 440 & 5  & 7,748 & 348    & 63 & 68,913 & 1,994 \\
    6     & 24  & 16,706 & 460 & 12  & 15,466 & 376 & 11  & 13,927 & 340 & 11  & 15,420 & 418 & 5  & 7,492 & 374    & 63 & 69,011 & 1,968 \\
    \bottomrule
  \end{tabular}
  }
  \label{tab:table1}
\end{table}

\begin{table}[ht] 
  \centering
    \caption{Scanning 3D U-Net architecture (K, O, I, D, W, and H are the kernel size, output channel, input channel, output depth, output width, and output height, respectively).}
      \scalebox{0.85}[0.85]{ 

  \begin{tabular}{crrrccc}
    \toprule
     Network Path & Layer & Size of K$_W$ × K$_H$ × O × I & Output Size (O × W × H) & Attention & Batch Normalization & Activation \\

    \midrule\midrule
         & conv1 & 3 × 3 × 64 × 1 & 64 × 256 × 256 & False & True & ReLU      \\
         & conv2 & 3 × 3 × 64 × 64 & 64 × 256 × 256 & False & True & ReLU     \\
         & conv3 & 3 × 3 × 128 × 64 & 128 × 128 × 128 & False & True & ReLU      \\
         & conv4 & 3 × 3 × 128 × 128 & 128 × 128 × 128 & False & True & ReLU    \\
Encoder  & conv5 & 3 × 3 × 256 × 128 & 256 × 64 × 64 & False & True & ReLU      \\
         & conv6 & 3 × 3 × 256 × 256 & 256 × 64 × 64 & False & True & ReLU      \\
         & conv7 & 3 × 3 × 512 × 256 & 512 × 32 × 32 & False & True & ReLU      \\
         & conv8 & 3 × 3 × 512 × 512 & 512 × 32 × 32 & False & True & ReLU      \\
         & conv9 & 3 × 3 × 1024 × 512 & 1024 × 16 × 16 & False & True & ReLU      \\
         & conv10 & 3 × 3 × 1024 × 1024 & 1024 × 16 × 16 & False & True & ReLU    \\

    \midrule
         & deconv1 & 3 × 3 × 512 × 1024 & 512 × 32 × 32 & True & True & ReLU      \\
         & deconv2 & 3 × 3 × 512 × 512 & 512 × 32 × 32 & True & True & ReLU      \\
         & deconv3 & 3 × 3 × 256 × 512 & 256 × 64 × 64 & True & True & ReLU      \\
         & deconv4 & 3 × 3 × 256 × 256 & 256 × 64 × 64 & True & True & ReLU     \\
Decoder  & deconv5 & 3 × 3 × 128 × 256 & 128 × 128 × 128 & True & True & ReLU      \\
         & deconv6 & 3 × 3 × 128 × 128 & 128 × 128 × 128 & True & True & ReLU     \\
         & deconv7 & 3 × 3 × 64 × 128 & 64 × 256 × 256 & True & True & ReLU    \\
         & deconv8 & 3 × 3 × 64 × 64 & 64 × 256 × 256 & True & True & ReLU      \\
         & conv & 1 × 1 × 6 × 64 & 6 × 256 × 256 & True & False & SoftMax      \\

    \bottomrule
  \end{tabular}
    }
  \label{tab:table2}
\end{table}

\begin{table}[ht]
 \caption{Inference accuracy of the
trained model. Evaluation of the inference results of the model for each cellular organelle, in each case of cross-validation and for each stage of the cell cycle, based on the IoU score (\ref{math:math1}).} 

\scalebox{0.75}[0.8]{ 

\begin{tabular}{cccccccccccc}
\midrule
Stage & \multicolumn{2}{c}{1} & \multicolumn{2}{c}{2} & \multicolumn{2}{c}{3} & \multicolumn{2}{c}{4} & \multicolumn{2}{c}{5} \\ \midrule
& 1 direction & 3 directions & 1 directions & 3 directions & 1 directions & 3 directions & 1 directions & 3 directions & 1 directions & 3 directions \\     \midrule\midrule

Cytoplasm & 0.869 & 0.882 & 0.790 & 0.830 & 0.783 & 0.806 & 0.786 & 0.796 & 0.778 & 0.804 \\ 
Plastid & 0.954 & 0.958 & 0.899 & 0.928 & 0.924 & 0.945 & 0.937 & 0.941 & 0.918 & 0.944 \\ 
Mitochondrion & 0.825 & 0.857 & 0.581 & 0.703 & 0.712 & 0.804 & 0.759 & 0.754 & 0.625 & 0.711 \\ 
Peroxisome & 0.548 & 0.535 & 0.516 & 0.596 & 0.483 & 0.601 & 0.509 & 0.546 & 0.148 & 0.135 \\ 
Nucleus & 0.852 & 0.883 & 0.734 & 0.801 & 0.705 & 0.697 & 0.701 & 0.696 & 0.597 & 0.630 \\ 
Background & 0.996 & 0.996 & 0.996 & 0.996 & 0.996 & 0.996 & 0.997 & 0.997 & 0.998 & 0.998 \\ 
\midrule
\end{tabular}
}
 
  \label{tab:table3}
\end{table}

\clearpage

\end{document}